\documentclass{webofc}
\usepackage[varg]{txfonts}   
\usepackage{subcaption}
\usepackage{xcolor}

\newcommand{\gev}{{\, \text{GeV}}}
\newcommand{\mev}{{\, \text{MeV}}}
\newcommand{\jpsi}{{J/\psi}}

\begin{document}
\title{The $Y(4230)$ as a $D_1 \bar{D}$ molecule}
%
%

\author{\firstname{Leon} \lastname{von Detten }\inst{1}\fnsep\thanks{\email{l.von.detten@fz-juelich.de}}
        \firstname{Christoph} \lastname{Hanhart}\inst{1}\fnsep\thanks{\email{c.hanhart@fz-juelich.de}} \and
        \firstname{Vadim} \lastname{Baru}\inst{2}\fnsep\thanks{\email{vadim.baru@tp2.rub.de}}
}

\institute{Institute for Advanced Simulation, Institut f\"ur Kernphysik and J\"ulich Center for Hadron Physics, Forschungszentrum J\"ulich, D-52425 J\"ulich, Germany 
\and
Institut f\"ur Theoretische Physik II, Ruhr-Universit\"at Bochum, D-44780 Bochum, Germany 
        }

\abstract{%
  We show that the currently available data are consistent with $Y(4230)$ being a $D_1 \bar{D}$ hadronic molecule. 
  By a simultaneous fit to data from $e^+ e^- \rightarrow D^0 D^{* -} \pi^+,\: \jpsi \pi^+ \pi^-,\: \jpsi K^+ K^-,\: h_c \pi^+ \pi^-,\: \jpsi \eta,\: \chi_{c0} \omega,\: \chi_{c1}(3872) \gamma$ and $\mu^+ \mu^-$,  we demonstrate 
  that this single state can explain the experimental signals in the mass range from $4.2$ to $4.35 \gev$. 
}
\maketitle
\section{Introduction}
\label{sec:intro}
In recent years a large number of states was found in the heavy quarkonium mass region that exhibit properties at odds with quark model predictions. For recent reviews see e.g.,
Refs.~\cite{Lebed:2016hpi,Esposito:2016noz,Olsen:2017bmm,Guo:2017jvc,Brambilla:2019esw,Chen:2022asf}. In this work we focus on the vector states in the energy range from $4.2$ to $4.35 \gev$. The current status of the analysis as given 
by the Review of Particle Physics 
by the PDG~\cite{ParticleDataGroup:2022pth} shows a large scatter of the resonance parameters for the $\psi(4230)$ aka $Y(4230)$, which were obtained by individual single channel analyses of different final states using one or multiple overlapping Breit-Wigner parameterizations. Moreover, an additional state labeled $Y(4320)$ is claimed by BESIII  to account for the highly asymmetric lineshape in the ${e^+ e^- \rightarrow \jpsi \pi \pi}$ cross section\cite{BESIII:2022qal}. However, it is located very close to the $D_1 \bar D$ threshold and not observed in any other final state of the $Y(4230)$.
We show that a single molecular $Y(4230)$,
interfering with the $\bar cc$ state $\psi(4160)$, naturally explains data in various channels.

\section{Molecular Considerations}
\label{sec:considerations}
\begin{figure}[h]
\centering
\includegraphics[width=.9\linewidth,clip]{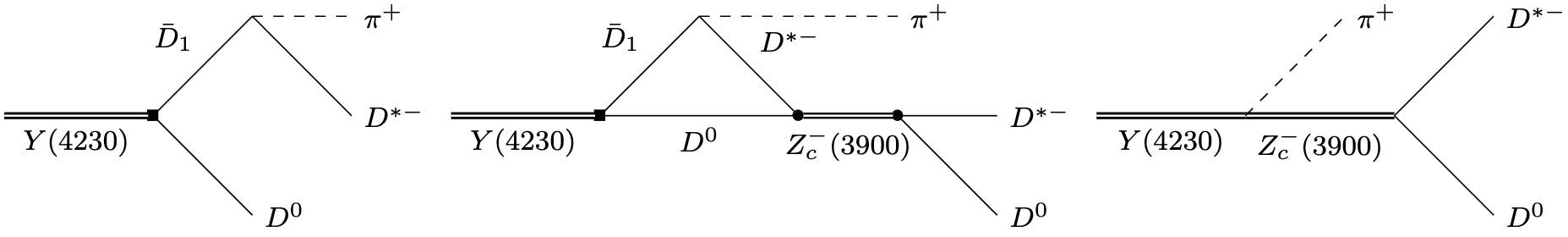}
\sidecaption
\caption{Leading diagrams contributing to $Y(4230)\rightarrow D^0 D^{* -} \pi^+$ in the molecular picture. }
\label{fig:feynman_R1} 
\end{figure}

\begin{figure}[h]
\centering
\includegraphics[width=\linewidth,clip]{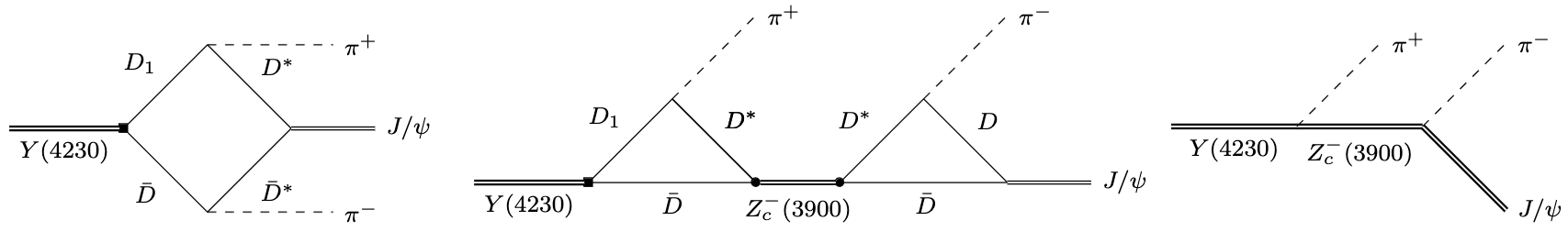}
\sidecaption
\caption{Main contributions to $Y(4230) \rightarrow \jpsi \pi^+ \pi^-$ in the molecular picture; only one representative topology is shown.}
\label{fig:feynman_R2} 
\end{figure}

If the $Y(4230)$ is a $D_1 \bar D$ molecule,
as proposed in Ref.~\cite{PhysRevLett.111.132003},
the Weinbergs compositeness criterium demands that the coupling of a molecular state to the two-hadron continuum that forms the molecule becomes maximal.
Therefore, loop contributions appear at leading order, resulting in potentially strong effects from the $D_1 \bar{D}$ threshold on different channels, as  demonstrated in Refs. \cite{Braaten:2007dw,Hanhart:2010wh}. 
In case of the $Y(4230)$ one of its constituents, the $D_1(2420)$, has a decay width of about $30 \mev$, decaying predominantly into $D^* \pi$. Hence the dominant decay mode with the largest connection to the molecular component is $D^* \bar{D} \pi$. This is  reflected in the experimental data, where $D^* \bar{D} \pi$ exhibits the largest cross section among the observed final states. The pertinent diagrams are shown in figure \ref{fig:feynman_R1}.
The $D^* \bar D$ as well as the $\jpsi \pi$ subsystem in $e^+ e^- \rightarrow \jpsi \pi \pi$ show clearly the $Z_c(3900)$, which in this work is treated as a molecular state of $D^* \bar{D}$, further enhancing the contribution of the triangle diagram. While the narrow $D_1(2420)$ in the heavy quark
limit decays to $D^*\pi$ in $d$-wave, the violation of heavy-quark spin symmetry (HQSS) may lead to its mixing with the much broader $D_1(2430)$,
where the $s$-wave decay is HQSS allowed. With its large width of over $300\mev$, the $D_1(2430)$ effectively acts as a short range interaction, allowing us to include its contribution via an s-wave point coupling of the $Y(4230)$ to $D^* \bar{D} \pi$. 
For the discovery channel of the $Y(4230)$, $e^+e^- \rightarrow \jpsi \pi \pi$, 
the leading order diagrams contain the $D_1 \bar{D}$ intermediate state via box and triangle topologies as shown in figure \ref{fig:feynman_R2}.

Additionally, in this channel the chiral Lagrangian allows for a direct point-like transition of $Y(4230)\rightarrow \jpsi \pi \pi$. The $\pi \pi/ \bar{K}K$ s-wave final state interaction (FSI) is included along the lines 
of the Muskhelishvili Omn\'es formalism outlined in, e.g.,~\cite{Chen:2019mgp,Baru:2020ywb,Danilkin:2020kce}. Here however,  we assume that  the principal value part of the Khuri-Treiman dispersion integral can be approximated by a contact polynomial (see $Y(4230)$ CT in Fig.~\ref{fig:fit_3body}).
For the 2-body final states $\jpsi \eta, \chi_{c0} \omega$ and $X(3872)\gamma$, the main contribution is given by a single triangle diagram, in combination with contact terms for HQSS allowed transitions~\cite{Baru:2022xne}.

For most final states the data call for an interference with an additional vector state --- thus we include the nearby $\psi(4160)$, which is
treated as a simple Breit-Wigner resonance, with the subsequent point-like decay of the $\psi(4160)$ to the given final state. 
\section{Results}
\begin{figure}
\centering
\begin{minipage}[t]{0.496\textwidth}
\vspace{-9.28cm}
\includegraphics[width=\linewidth,trim={30pt 15pt 5pt 5pt}]{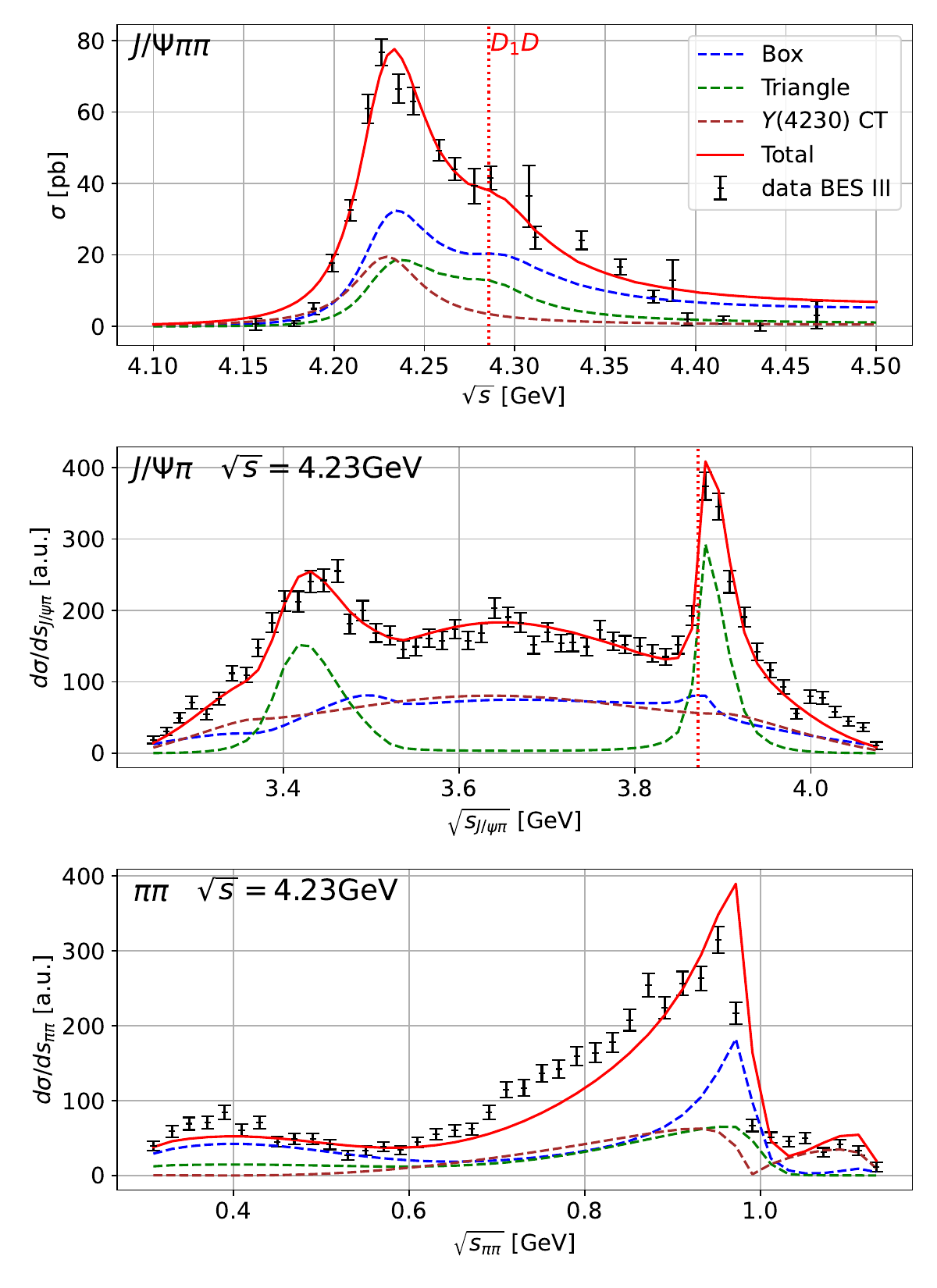}
\end{minipage}
\begin{minipage}[t]{0.496\textwidth}
\includegraphics[width=\linewidth,trim={30pt 5pt 5pt 5pt}]{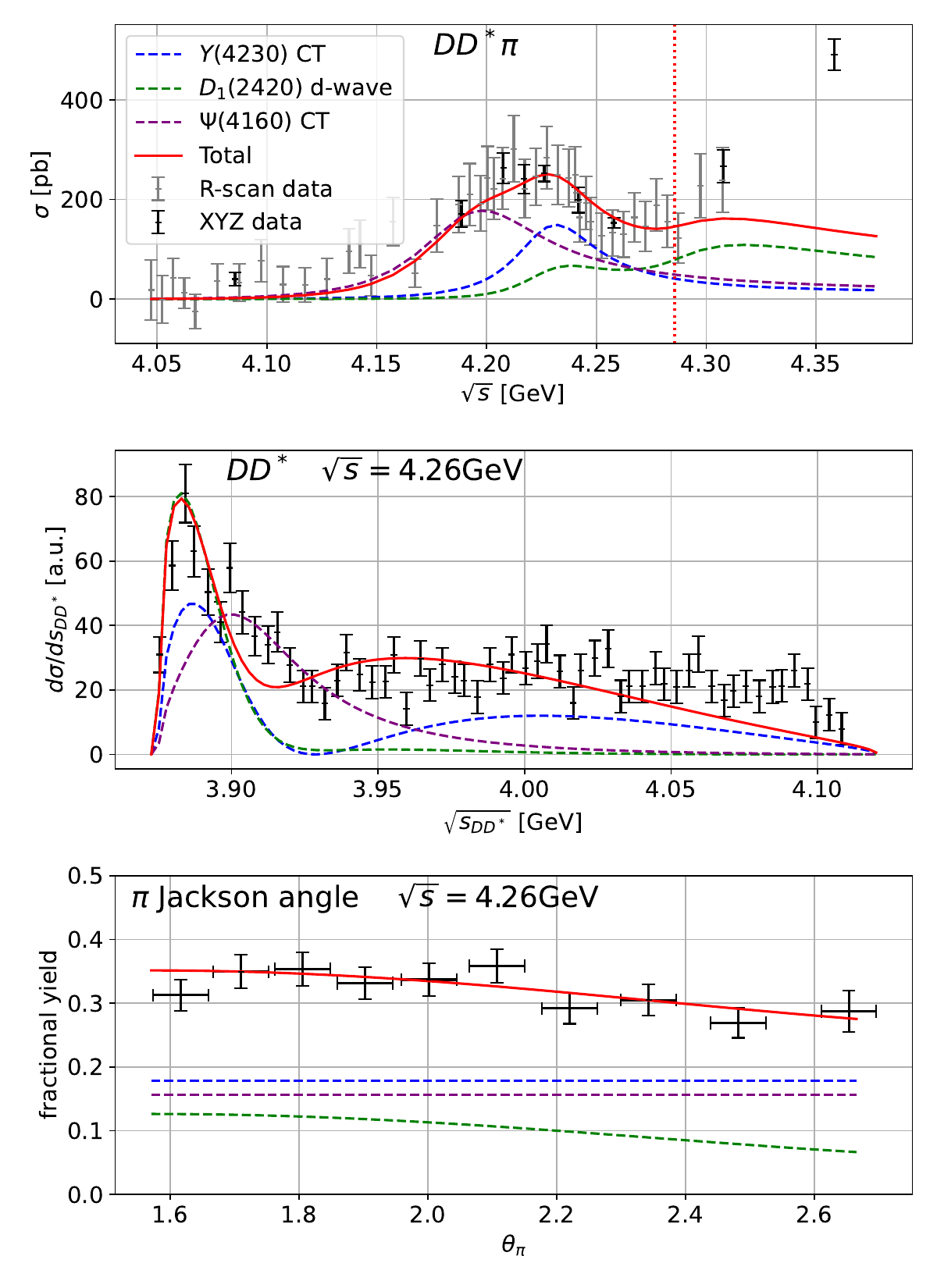}
\end{minipage}
\vspace{-.5cm}
\sidecaption
\caption{The left column shows the results for the $\jpsi \pi \pi$ cross section and the invariant mass spectra of the $\jpsi\pi$ and $\pi\pi$  subsystems,
in order. The right column shows the $D^0 D^{* -} \pi^+$ cross section, the $D \bar D^*$ subsystem and the pion Jackson angle. The data are taken from Refs.~\cite{BESIII:2022qal,BESIII:2017bua,BESIII:2023cmv,BESIII:2015pqw}.}
\label{fig:fit_3body}
\end{figure}
\begin{figure}
\centering
\begin{minipage}[t]{0.496\textwidth}
\includegraphics[width=\linewidth,trim={30pt 5pt 5pt 5pt}]{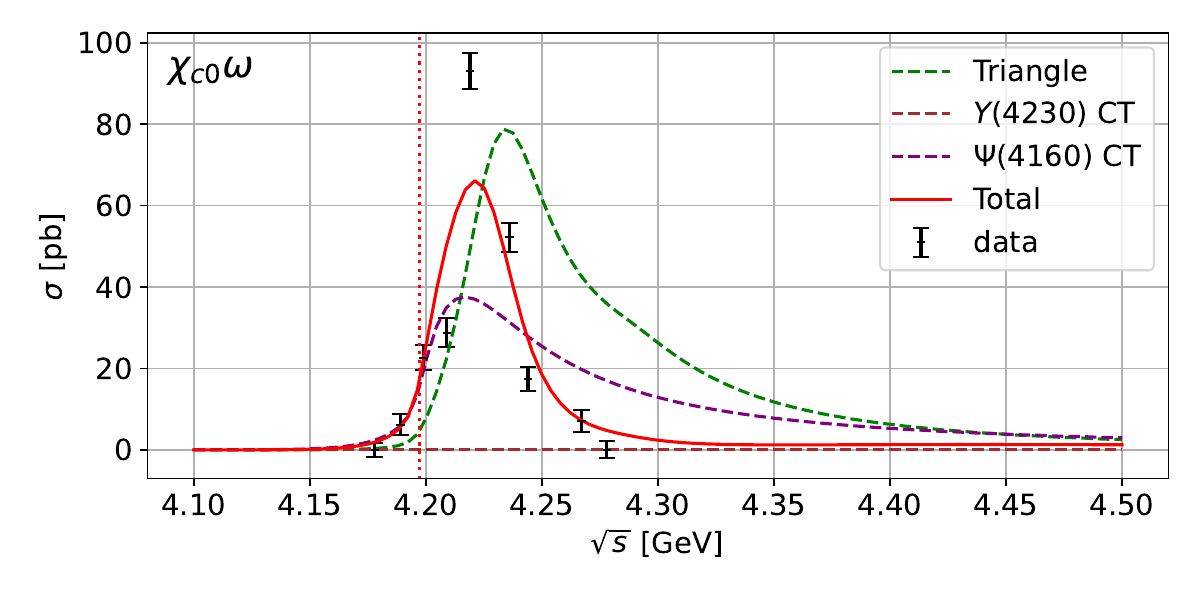}
\includegraphics[width=\linewidth,trim={30pt 5pt 5pt 5pt}]{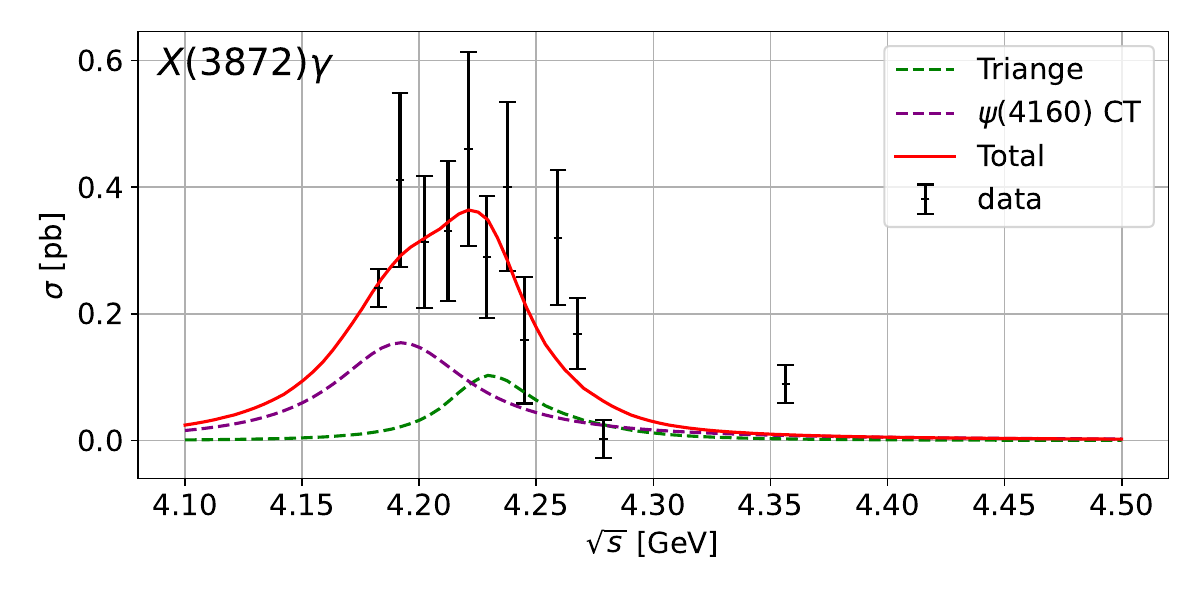}
\end{minipage}
\begin{minipage}[t]{0.496\textwidth}
\includegraphics[width=\linewidth,trim={30pt 5pt 5pt 5pt}]{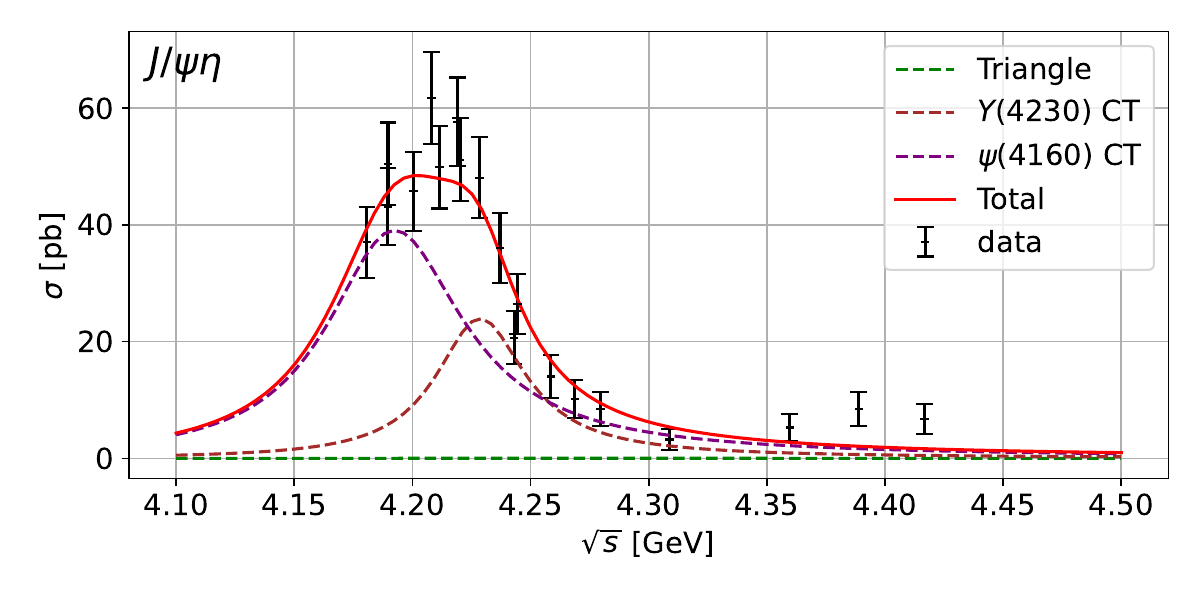}
\includegraphics[width=\linewidth,trim={30pt 5pt 5pt 5pt}]{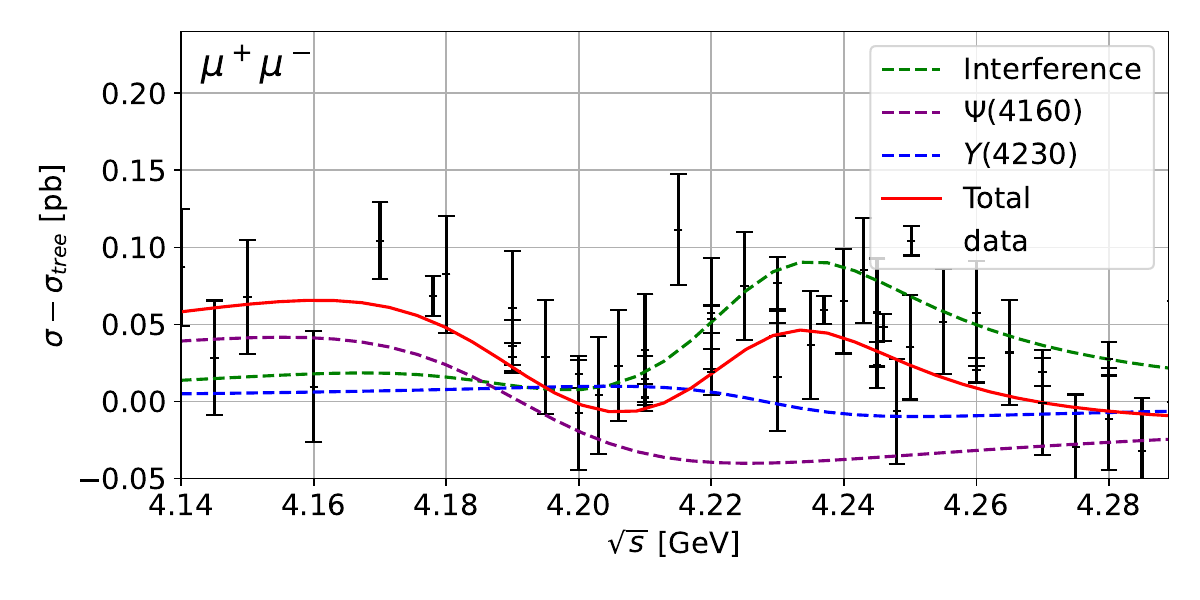}
\end{minipage}
\vspace{-.1cm}
\sidecaption
\caption{Fit results for the 2-body final states $\chi_{c0} \omega,\: \jpsi \eta,\: X(3872) \gamma$\: and $\mu^+ \mu^-$. The data are taken from Refs. \cite{BESIII:2019gjc,BESIII:2020bgb,BESIII:2019qvy,BESIII:2020peo}.}
\label{fig:fit_2body}
\end{figure}
The results for the $\jpsi \pi^+ \pi^-$ channel are shown in the left column of figure \ref{fig:fit_3body}. 
In the molecular picture, 
dominant loop contributions from the box and triangle diagrams enhance the cross section at the $D_1 \bar D$ threshold, allowing for the description of the asymmetric lineshape with just a single pole. The $\jpsi \pi$ subsystem shows a prominent peak from the $Z_c(3900)$ and its reflection, while the box and contact term act as a background. The general features of the $\pi \pi$ lineshape are also well reproduced, resulting from an interplay of various mechanisms. The slight deviation from the experimental data most likely arises from the approximate treatment of the $\pi \pi$ FSI. Moving to $D^0 D^{* -} \pi^+$ data, we found that it is not possible to describe the new high statistics data simultaneously with $\jpsi \pi^+ \pi^-$ including only the $Y(4230)$, as the left flank of peak falls off significantly faster for $\jpsi \pi^+ \pi^-$ 
than for $D^0 D^{* -} \pi^+$. However, the inclusion of the close by $\psi(4160)$ with a mass of $4191\pm5 \mev$ solves this problem. Furthermore, the interplay of the $Y(4230)$ pole with the $D_1 \bar D$ continuum provides a non-trivial $s$-dependence for the cross section. At higher energies the fitted lineshape shows a strong deviation from the data,
which are, however, expected:
$D_1 \bar D^*$ and $D_2 \bar D^*$ can also form bound states in the molecular scenario~\cite{Wang:2013kra,Ji:2022blw}. The $D\bar D^*$ lineshape deviates slightly from the data, but the Jackson angle of the pion is well reproduced. 
We do not show our results for the remaining three-body final states here, but only discuss them briefly:
The contributions to the $h_c \pi^+ \pi^-$ channel are similar to those in $\jpsi \pi^+ \pi^-$ shown in Fig.~\ref{fig:feynman_R2}, except for the contact term, which is subleading due to HQSS violation.
-- see also the discussion in~\cite{Baru:2022xne}. The $h_c \pi$ subsystem shows a very prominent peak of the $Z_c(4020)$, requiring the couplings of the $Y(4230)$ to $D_1 \bar D^*$ or/and the $Z_c(4020)$ to $D \bar D^*$, which are omitted for now. 
On the other hand,  the box diagrams are again able to describe the gross features of this lineshape, showing again a big imprint of the $D_1 \bar D$ threshold. We treat this as a  consistency check.
Similarly, the consistent treatment of 
 the $\jpsi K^+ K^-$ final state would require the $Z_{cs}(4000)$ to appears in the triangle (see, e.g., Refs~\cite{Baru:2021ddn,Ortega:2021enc,Du:2022jjv} for the molecular interpretation of this state based on data), whose inclusion is postponed for the full coupled channel analysis. However, apart from the missing coupling strength, the box and the contact term are already able to reproduce the $\jpsi K^+ K^-$ data. The data for the remaining 2 body final states also require the inclusion of the $\psi(4160)$ in combination with the $Y(4230)$ in either a triangle diagram or contact term --- the results are shown in Fig.~\ref{fig:fit_2body}.
The very narrow peak in $\chi_{c0} \omega$ is 
explained via a destructive interference between the two resonances. $\jpsi \eta$ and $X(3872) \gamma$ are well described. Even though $\psi(4160)$ and $Y(4230)$ contributions to the hadronic cross sections discussed above are of comparable order, the production rate from $e^+ e^- \rightarrow \gamma^*$ is larger for the $\psi(4160)$ in comparison to that of the $Y(4230)$, since the latter in the molecular scenario is suppressed by HQSS.
\begin{figure}[h]
\centering
\includegraphics[width=.9\linewidth,clip]{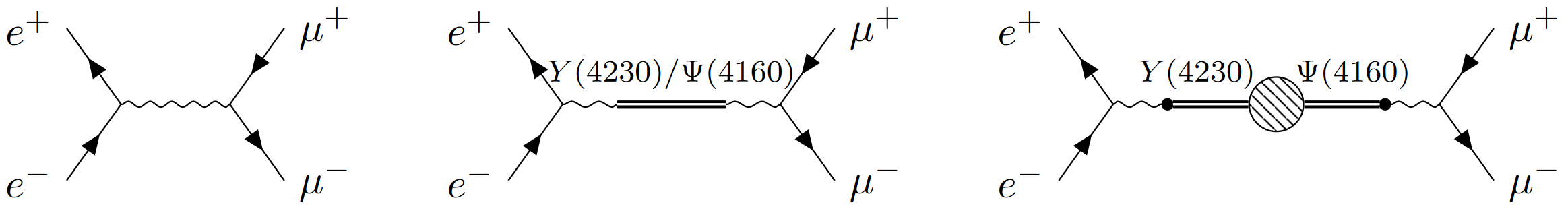}
\sidecaption
\caption{Contributions for $e^+ e^-\rightarrow \mu^+ \mu^-$. }
\label{fig:feynman_R9} 
\end{figure}
This is consistent with our fit to $\mu^+ \mu^-$ (shown in Figure \ref{fig:fit_2body}), where the contribution of the $\psi(4160)$, rescaled by the squared resonance-to-photon coupling, dominates over the $Y(4230)$.
 Furthermore, the fit shows clearly an interference between the two resonances given by the last diagram in Fig.~\ref{fig:feynman_R9}.  For this calculation we absorb the full real part 
 of the pertinent transition matrix element
 into a constant adjusted to the data, while its imaginary part  is completely fixed by 
 unitarity through  the contributions discussed above. We find that
the fit calls for the latter of the same order as the former, as expected from naturalness. Thus the good description of the $\mu^+\mu^-$ data is a very non-trivial consistency check of the analysis outlined above.

In conclusion, the molecular scenario for the $Y(4230)$ together with an interference with the $\psi(4160)$, nicely reproduces the experimental lineshapes of various final states
with the quantum numbers $J^{PC}=1^{--}$  
in the mass range from $4.2$ to $4.35 \gev$.\medskip\\

We thank Ryan Mitchell for the idea to include the $\psi(4160)$ in the analysis.
This work is supported in part by the
Deutsche Forschungsgemeinschaft (DFG) through the funds provided to the Sino-German Collaborative Research Center TRR110 ``Symmetries and the Emergence of Structure in QCD'' (NSFC Grant No. 12070131001, DFG Project-ID 196253076),
by the NSFC under Grants No.~11835015, No.~12047503, No.~11961141012, and No.~12035007, and by the Chinese Academy of Sciences under Grants
No. QYZDB-SSW-SYS013, No. XDB34030000, No.~XDPB15 and No.~2020VMA0024.

\bibliography{refs.bib}

\begin{thebibliography}{27}

\bibitem{Lebed:2016hpi}
R.F. Lebed, R.E. Mitchell, E.S. Swanson, Prog. Part. Nucl. Phys. \textbf{93},
  143 (2017), \texttt{1610.04528}

\bibitem{Esposito:2016noz}
A.~Esposito, A.~Pilloni, A.D. Polosa, Phys. Rept. \textbf{668}, 1 (2017),
  \texttt{1611.07920}

\bibitem{Olsen:2017bmm}
S.L. Olsen, T.~Skwarnicki, D.~Zieminska, Rev. Mod. Phys. \textbf{90}, 015003
  (2018), \texttt{1708.04012}

\bibitem{Guo:2017jvc}
F.K. Guo, C.~Hanhart, U.G. Mei\ss{}ner, Q.~Wang, Q.~Zhao, B.S. Zou, Rev. Mod.
  Phys. \textbf{90}, 015004 (2018), \texttt{1705.00141}

\bibitem{Brambilla:2019esw}
N.~Brambilla, S.~Eidelman, C.~Hanhart, A.~Nefediev, C.P. Shen, C.E. Thomas,
  A.~Vairo, C.Z. Yuan, Phys. Rept. \textbf{873}, 1 (2020), \texttt{1907.07583}

\bibitem{Chen:2022asf}
H.X. Chen, W.~Chen, X.~Liu, Y.R. Liu, S.L. Zhu, Rept. Prog. Phys. \textbf{86},
  026201 (2023), \texttt{2204.02649}

\bibitem{ParticleDataGroup:2022pth}
R.L. Workman et~al. (Particle Data Group), PTEP \textbf{2022}, 083C01 (2022)

\bibitem{BESIII:2022qal}
M.~Ablikim et~al. (BESIII), Phys. Rev. D \textbf{106}, 072001 (2022),
  \texttt{2206.08554}

\bibitem{PhysRevLett.111.132003}
Q.~Wang, C.~Hanhart, Q.~Zhao, Phys. Rev. Lett. \textbf{111}, 132003 (2013)

\bibitem{Braaten:2007dw}
E.~Braaten, M.~Lu, Phys. Rev. D \textbf{76}, 094028 (2007), \texttt{0709.2697}

\bibitem{Hanhart:2010wh}
C.~Hanhart, Y.S. Kalashnikova, A.V. Nefediev, Phys. Rev. D \textbf{81}, 094028
  (2010), \texttt{1002.4097}

\bibitem{Chen:2019mgp}
Y.H. Chen, L.Y. Dai, F.K. Guo, B.~Kubis, Phys. Rev. D \textbf{99}, 074016
  (2019), \texttt{1902.10957}

\bibitem{Baru:2020ywb}
V.~Baru, E.~Epelbaum, A.A. Filin, C.~Hanhart, R.V. Mizuk, A.V. Nefediev,
  S.~Ropertz, Phys. Rev. D \textbf{103}, 034016 (2021), \texttt{2012.05034}

\bibitem{Danilkin:2020kce}
I.~Danilkin, D.A.S. Molnar, M.~Vanderhaeghen, Phys. Rev. D \textbf{102}, 016019
  (2020), \texttt{2004.13499}

\bibitem{Baru:2022xne}
V.~Baru, E.~Epelbaum, A.A. Filin, C.~Hanhart, A.V. Nefediev, Phys. Rev. D
  \textbf{107}, 014027 (2023), \texttt{2211.08038}

\bibitem{BESIII:2017bua}
M.~Ablikim et~al. (BESIII), Phys. Rev. Lett. \textbf{119}, 072001 (2017),
  \texttt{1706.04100}

\bibitem{BESIII:2023cmv}
M.~Ablikim et~al. (BESIII), Phys. Rev. Lett. \textbf{130}, 121901 (2023),
  \texttt{2301.07321}

\bibitem{BESIII:2015pqw}
M.~Ablikim et~al. (BESIII), Phys. Rev. D \textbf{92}, 092006 (2015),
  \texttt{1509.01398}

\bibitem{BESIII:2019gjc}
M.~Ablikim et~al. (BESIII), Phys. Rev. D \textbf{99}, 091103 (2019),
  \texttt{1903.02359}

\bibitem{BESIII:2020bgb}
M.~Ablikim et~al. (BESIII), Phys. Rev. D \textbf{102}, 031101 (2020),
  \texttt{2003.03705}

\bibitem{BESIII:2019qvy}
M.~Ablikim et~al. (BESIII), Phys. Rev. Lett. \textbf{122}, 232002 (2019),
  \texttt{1903.04695}

\bibitem{BESIII:2020peo}
Phys. Rev. D \textbf{102}, 112009 (2020), \texttt{2007.12872}

\bibitem{Wang:2013kra}
Q.~Wang, M.~Cleven, F.K. Guo, C.~Hanhart, U.G. Mei\ss{}ner, X.G. Wu, Q.~Zhao,
  Phys. Rev. D \textbf{89}, 034001 (2014), \texttt{1309.4303}

\bibitem{Ji:2022blw}
T.~Ji, X.K. Dong, F.K. Guo, B.S. Zou, Phys. Rev. Lett. \textbf{129}, 102002
  (2022), \texttt{2205.10994}

\bibitem{Baru:2021ddn}
V.~Baru, E.~Epelbaum, A.A. Filin, C.~Hanhart, A.V. Nefediev, Phys. Rev. D
  \textbf{105}, 034014 (2022), \texttt{2110.00398}

\bibitem{Ortega:2021enc}
P.G. Ortega, D.R. Entem, F.~Fernandez, Phys. Lett. B \textbf{818}, 136382
  (2021), \texttt{2103.07871}

\bibitem{Du:2022jjv}
M.L. Du, M.~Albaladejo, F.K. Guo, J.~Nieves, Phys. Rev. D \textbf{105}, 074018
  (2022), \texttt{2201.08253}

\end{thebibliography}

\end{document}